\documentclass[3p,twocolumn,authoryear]{elsarticle}

\usepackage{makecell}
\usepackage[T1]{fontenc}
\usepackage[english]{babel}		
\usepackage[utf8]{inputenc}
\usepackage{textcomp}
\usepackage{amsmath}
\usepackage{graphicx}
\usepackage{systeme}
\usepackage{multirow}
\usepackage{geometry}
\usepackage{natbib}
\usepackage{adjustbox}
\usepackage{xcolor}
\usepackage{mathtools}
\usepackage{subfig}
\usepackage{floatrow}
\usepackage{url}
\usepackage{epstopdf}
\usepackage{url}
\usepackage{subfig}
\usepackage{amssymb}
\usepackage{etoolbox}
\usepackage{amsthm}
\usepackage{booktabs}
\usepackage{algorithm}
\usepackage[noend]{algpseudocode}
\usepackage[algo2e, ruled, lined, linesnumbered]{algorithm2e}

\newtheorem*{remark}{Remark}

\newcommand{\ubar}[1] {\underline{\smash{#1}}}
\newcommand{\obar}[1] {\overline{#1}}

\newcommand{\nonl}{\renewcommand{\nl}{\let\nl\oldnl}} 
\newcommand{\argminF}{\mathop{\mathrm{argmin}}\limits}   
\DeclareMathOperator*{\argmin}{arg\,min}

\def\@author#1{\g@addto@macro\elsauthors{\normalsize%
		\def\baselinestretch{1}%
		\upshape\authorsep#1\unskip\textsuperscript{%
			\ifx\@fnmark\@empty\else\unskip\sep\@fnmark\let\sep=,\fi
			\ifx\@corref\@empty\else\unskip\sep\@corref\let\sep=,\fi
		}%
		\def\authorsep{\unskip,\space}%
		\global\let\@fnmark\@empty
		\global\let\@corref\@empty  
		\global\let\sep\@empty}%
	\@eadauthor={#1}
}

\theoremstyle{definition}

\theoremstyle{remark}

\begin{document}
\begin{frontmatter}

\title{Optimization tools for Twin-in-the-Loop vehicle control design:\\ analysis and yaw-rate tracking case study}

\author[a]{Federico Dettù}
\ead{federico.dettu@polimi.it}
\author[a]{Simone Formentin\corref{cor1}}
\ead{simone.formentin@polimi.it}
\author[b]{Stefano Varisco} 
\ead{stefano.varisco@ferrari.com}
\author[a]{Sergio Matteo Savaresi} 
\ead{sergio.savaresi@polimi.it}
\address[a]{Dipartimento di Elettronica, Informazione e Bioingegneria, Politecnico di Milano. \\Via G. Ponzio 34/5, 20133, Milano, Italy.}
\address[b]{Vehicle Dynamics, Ferrari S.p.A. \\Via Abetone Inferiore 4, 41053, Maranello, Italy.}
\cortext[cor1]{Corresponding author.}

\begin{abstract}
Given the urgent need of simplifying the end-of-line tuning of complex vehicle dynamics controllers, the Twin-in-the-Loop Control (TiL-C) approach was recently proposed in the automotive field. In TiL-C, a digital twin is run on-board to compute a nominal control action in run-time and an additional block $C_{\delta}$ is used to compensate for the mismatch between the simulator and the real vehicle. As the digital twin is assumed to be the best replica available of the real plant, the key issue in TiL-C becomes the tuning of the compensator, which must be performed relying on data only. In this paper, we investigate the use of different black-box optimization techniques for the calibration of $C_{\delta}$. More specifically, we compare the originally proposed Bayesian Optimization (BO) approach with the recently developed Set Membership Global Optimization (SMGO) and Virtual Reference Feedback Tuning (VRFT), a one-shot direct data-driven design method. The analysis will be carried out within a professional multibody simulation environment on a novel TiL-C application case study -- the yaw-rate tracking problem -- so as to further prove the TiL-C effictiveness on a challenging problem. Simulations will show that the VRFT approach is capable of providing a well tuned controller after a single iteration, while $10$ to $15$ iterations are necessary for refining it with global optimizers. Also, SMGO is shown to significantly reduce the computational effort required by BO.
\end{abstract}

\end{frontmatter}

\section{Introduction}
\label{Section:til_opt_Introduction}
\begin{figure}[h]
	\centering
	\includegraphics[width=0.8 \columnwidth]{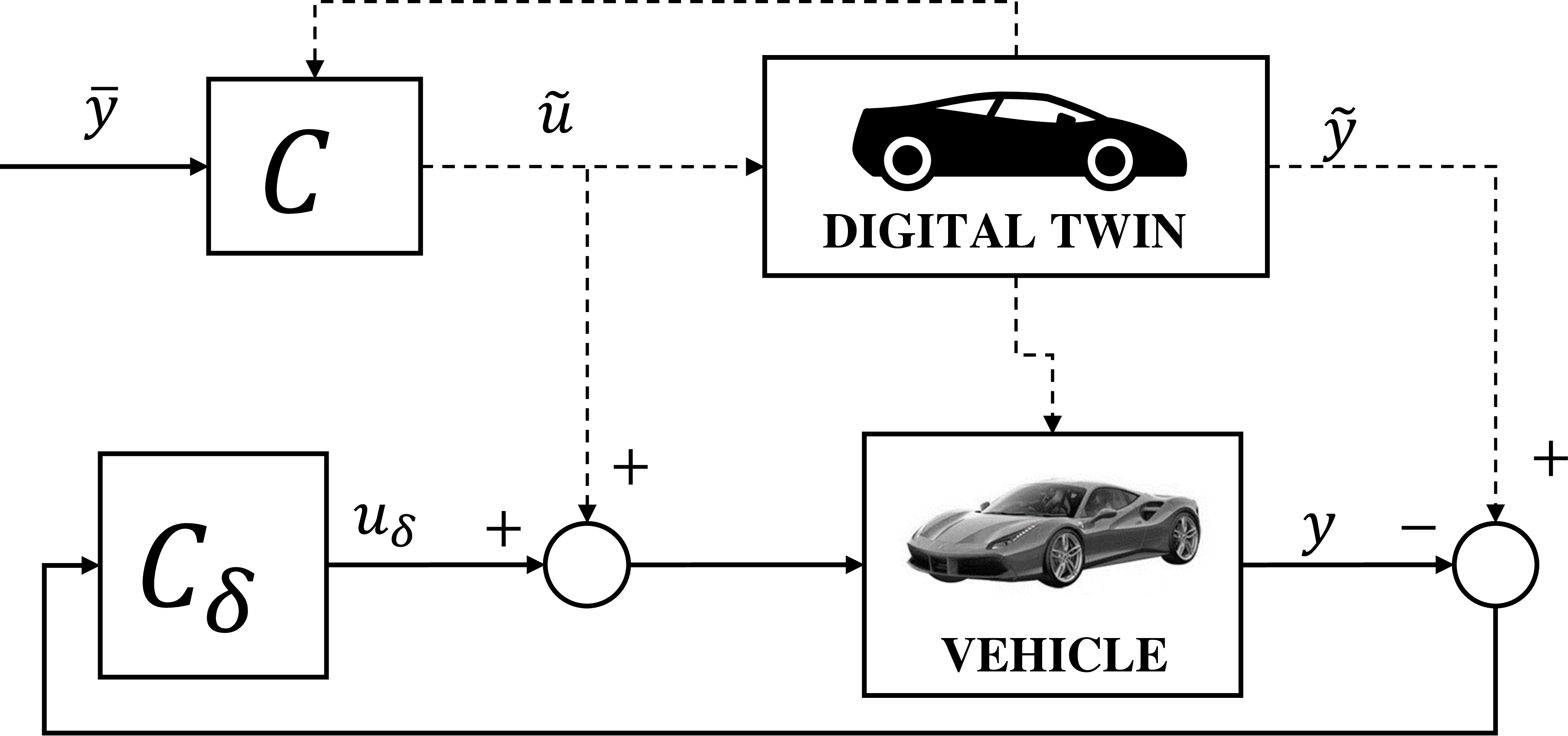}
	\caption{Twin-in-the-Loop vehicle control scheme.}
	\label{fig:til_opt_sil_c_general}
\end{figure}
Although control theorists have been pushing forward for decades in developing well performing and sophisticate control algorithms, a sort of detachment arose when looking at what control practicioners do. A few years ago, the survey \citep{samad2017survey} showed that Proportional Integral Derivative control is perceived as the most impacting technology in industry, far above \emph{e.g.} nonlinear or adaptive controls, and somewhat more relevant than extremely popular Model Predictive Control. In the same survey, it emerges that the lack of competences in industry is an important perceived issue when dealing with modern controllers. In fact, calibrating and updating a developed controller requires adjustment of several tuning knobs, whose effect is often poorly understood without a deep knowledge of control theory.

Concentrating our efforts toward this direction, we recently proposed an approach for simplifying the implementation of complex controllers and easing the end-of-line tuning phase in the vehicle domain: the Twin-in-the-Loop Control (TiL-C) framework\footnote{The dual problem of Twin-in-the-Loop Filtering (TiL-F) is also under investigation \citep{riva_2022_SIL,delcaro_2023}.} \citep{dettu2023twin}, schematically represented in Fig. \ref{fig:til_opt_sil_c_general}\footnote{A discussion on robust calibration for TiL controllers has been proposed in \cite{dettu2023rob}.}.
TiL-C runs a digital twin (DT) --a highly sophisticated model-- inside a vehicle control unit in real-time. This is made possible by the most recent technological advances (see \emph{e.g.} the Autohawk platform \cite{autohawk}). Further, in the automotive domain, DT often come for free, as car manufacturer already have a portfolio of well calibrated models of their production vehicles, used for off-line testing and virtual prototyping. \\ The TiL-C architecture is based upon a nominal controller $C$; such a controller is calibrated on the DT and then used on-board at run-time. This allows for the computation of a nominal control action to be applied \textit{as a first -  open-loop - contribution} to the real vehicle. It was then shown that a simple compensator $C_{\delta}$ (\emph{e.g.} a PID) could be sufficient to guarantee stability and maintain the same performance attained on the DT \textit{also on the real vehicle}. By using TiL-C, the end-of-line tuning shifts from the implementation (and sometimes redesign) of a complex controller, to the tuning of a few gains, which is more easily done and widely understood.

An important issue in TiL-C -- deeply discussed in \citep{dettu2023twin} -- is how to use data to calibrate such parameters of the compensator $C_{\delta}$: since this controller manages the unknown dynamics between the DT and the vehicle, no classical model-based design techniques (\emph{e.g.} loop shaping) can be used. Following a recently established literature trend \citep{dettu2023modeling,busetto2023data,coutinho2023bayesian}, we first proposed Bayesian Optimization (BO) to solve this problem. Although effectiveness of BO in calibrating TiL-C has been shown ($100$ iterations have been considered), no discussions concerning the selection of a specific optimization algorithm have been considered. Also, no analyses concerning the number of iterations necessary to get to convergence has been carried out. Being vehicle testing very costly, assessing (and possibly reducing) the number of the necessary experiments is extremely important.
We thus concentrate our efforts to investigate and analyse different approaches for calibrating a TiL controller.

\textbf{Related works.} Within the context of black-box optimization (of which BO is a particular case), many approaches have been developed throughout the years and successfully applied to different fields.
Two classic iterative methods are \emph{e.g.} Extremum Seeking (ES) \citep{hellstrom2013board} or Iterative Feedback Tuning (IFT) \citep{hjalmarsson1998iterative}. Although being simple to apply, these approaches lost popularity, suffering from different issues, such as the high number of iterations required to get a solution, or the impossibility of including safety a-priori unknown constraints during the tuning process, potentially yielding hazardous controller attempts. Furthermore, these methods are not \textit{global}, and they might become stuck in a local minimum, requiring re-initialization and thus increased costs.

Among \textit{global} optimizers, surrogate function based ones (such as BO) gained much popularity with respect to \emph{e.g.} genetic algorithms \citep{jaen2013pid} or particle swarm \citep{kim2008robust} ones, due to the large --often unfeasible in real applications-- number of iterations required by the latter.
As already mentioned, BO \citep{shahriari2015taking} proved to be very effective when solving black-box optimization problems: calibrating a controller for a partially or totally unknown plant lies within this framework. BO fits a Gaussian Process (GP) surrogate of the unknown cost function to be minimized, and iteratively selects new candidate points to be evaluated based on the acquired GP knowledge. 
A nice BO feature is the possibility of including a-priori unknown constraints within the optimization problems: similarly as done for the cost function, these constraint functions are iteratively estimated from data. This opens up the possibility of \emph{e.g.} maintaining the system in a safe region throughout the calibration procedure.

The most significant issue with BO is the high computational burden yield by the GP fitting and optimization: the more data points are added, the more cumbersome this procedure becomes, undermining application of BO in low-cost processing units. To solve this, \cite{sabug2022smgo} recently presented Set Membership Global Optimization-$\Delta$ (SMGO-$\Delta$); with respect to BO, this approach provides a computationally efficient objective and constraints model, based on hypercones, and focusing on quantifying the uncertainty associated to the unknown function, rather than directly trying to fit it. SMGO-$\Delta$ also allows accounting for black-box constraints, and the authors proven its convergence under the assumption of Lipschitz continuity of the objective and constraints. SMGO-$\Delta$ has recently been used in practical control tuning problems \citep{galbiati2022direct,busetto2023data,sabug2023simultaneous}.

A completely different philosophy when dealing with controller tuning is the Virtual Reference Feedback Tuning (VRFT) one: VRFT is \textit{direct} data-driven approach, thus requiring a single experiment for learning the controller. It is based upon frequency-domain considerations, and tries to match a reference closed-loop behaviour. VRFT has been successfully used in different control tuning problems, \emph{e.g.} electric motors \citep{busetto2023data}, engines \citep{passenbrunner2014direct} or brake-by-wire systems \citep{radrizzani2020data}; however, some critical tuning knobs (like the reference model) exist when applying VRFT, and the resulting controller might be underperforming or even unstable if they are not properly selected. \\ 
Finally, although well promising, TiL-C has been so far exclusively applied to braking control \citep{dettu2023rob,dettu2023twin}: in the strive of widening the portfolio of applications and further proofing its efficacy, different case studies should be considered.

\textbf{Contributions.} Taking into account considerations above, we consider here SMGO-$\Delta$ for TiL calibration, comparing it with BO, and discussing the convergence speed of both algorithms. We include a constraint during the optimization process, in order to guarantee that unsafe regions are avoided. Also, we apply VRFT to the same problem, eventually showing the significant performance obtained with just one experiment. Combination of both methods (VRFT and BO/SMGO) is also explored.
On top of this, we apply TiL-C to the vehicle yaw-rate tracking problem: such a problem is important both in production vehicles driving assistance systems -- it is at the basics of Electronic Stability Control (ESC) \cite{gimondi2021mixed} -- and in autonomous driving, where it is combined with trajectory following \citep{corno2023optimization,falcone2007predictive}. Being an established state-of-the-art \citep{lucchini2020torque,lucchini2021design,spielberg_2022,beal2012model,falcone2007predictive}, we use Model Predictive Control as a baseline controller in yaw-rate tracking.
Yaw-rate control poses critical stability problems, in that the vehicle sideslip might utterly increase if not properly addressed, leading to unsafe situations and undesired drifting of the vehicle (\emph{i.e.} too large wheel slips): it is thus an interesting and challenging test bench for TiL-C.\\

The remainder of the paper is as follows. In Section \ref{Section:til_opt_nominal_controller}, we describe the yaw-rate control problem, the simplified model of the vehicle and derive the baseline controller, to be used within TiL-C. Then, in Section \ref{Section:til_opt_til_control}, we show how TiL-C can be applied to the presented problem, and discuss the different optimization tools. Section \ref{Section:til_opt_simulation_results} compares the optimizers in terms of cost function minimization and constraints violation, and then gives an in-depth time-domain analysis of the obtained results, on different maneuvers. 

\section{The yaw-rate tracking problem}
\label{Section:til_opt_nominal_controller}
The yaw-rate tracking problem is tackled in this research with Twin-in-the-Loop control. This application problem is used to evaluate and analyse different optimization tools for TiL-C. Hence, we start off by precisely defining a control-oriented system model and a baseline controller. The goal is that of following a reference yaw-rate $r_{ref}$ -- generated by the driver steering action $s_{ref}$ by means of a suitable reference generation mechanism. The controller decides a steering action $s_{cmd}$ to be passed to a steer-by-wire system, acting onto the front wheels. Measurements used for feedback control are: the yaw-rate $r$, the vehicle sideslip angle $\beta$ (usually retrieved by means of an estimation algorithm \citep{carnier2023hybrid}, not in the scopes of this work), the steering actuator position $s_{act}$, the longitudinal component of vehicle speed $v_x$ and the longitudinal acceleration $a_x$.

\subsection{System modelling}
\label{Section:til_opt_control_oriented_model}

In the following, we describe the model used for developing lateral dynamic controllers. The model takes into account simplified vehicle dynamics and the steer-by-wire actuator.
\paragraph{Vehicle model}
\begin{figure}[h]
	\centering
	\includegraphics[width=0.8 \columnwidth]{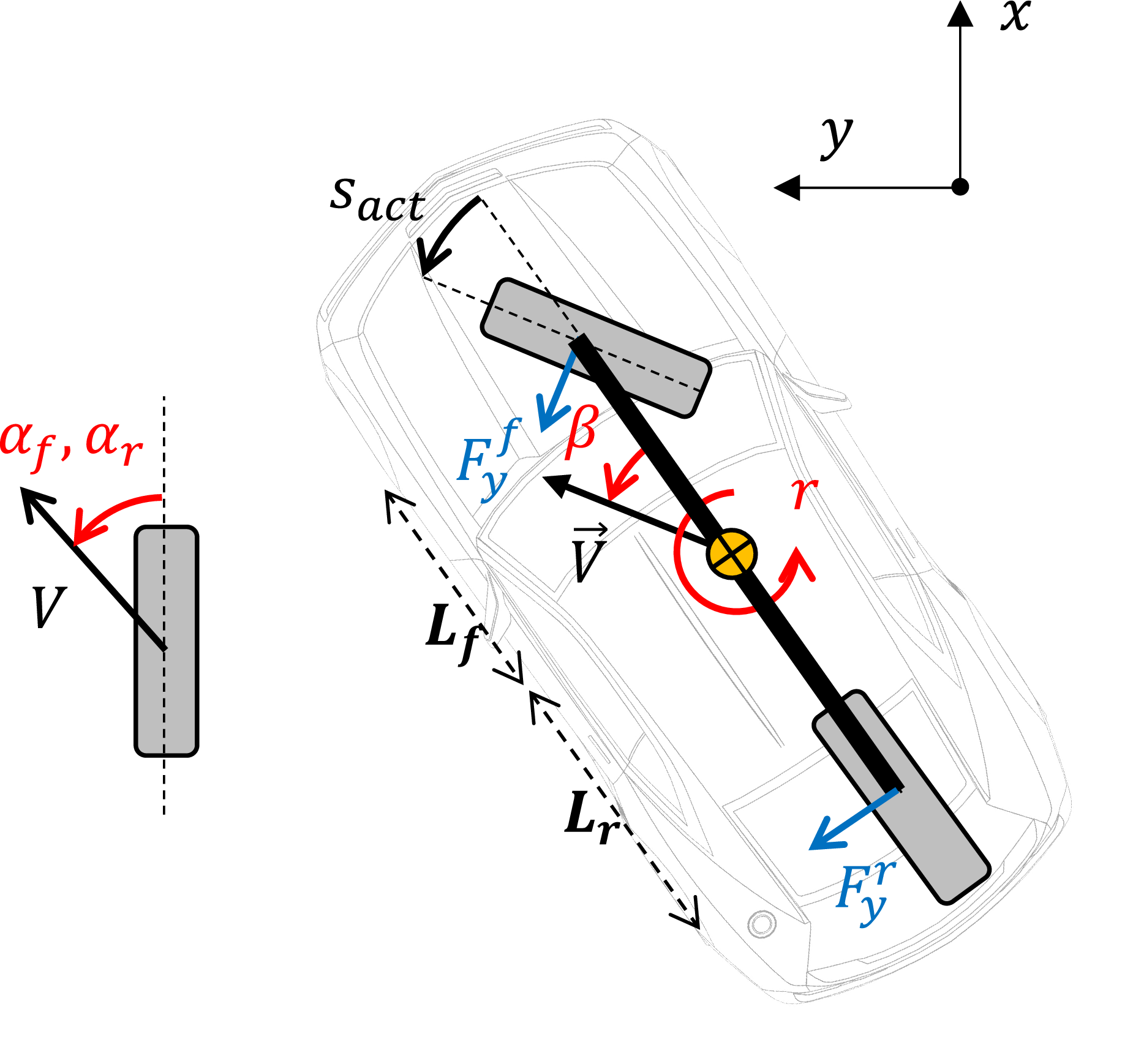}
	\caption{Single-track vehicle model.}
	\label{fig:til_opt_single_track_model}
\end{figure}
\begin{figure}[h]
	\centering
	\includegraphics[width=0.8 \columnwidth]{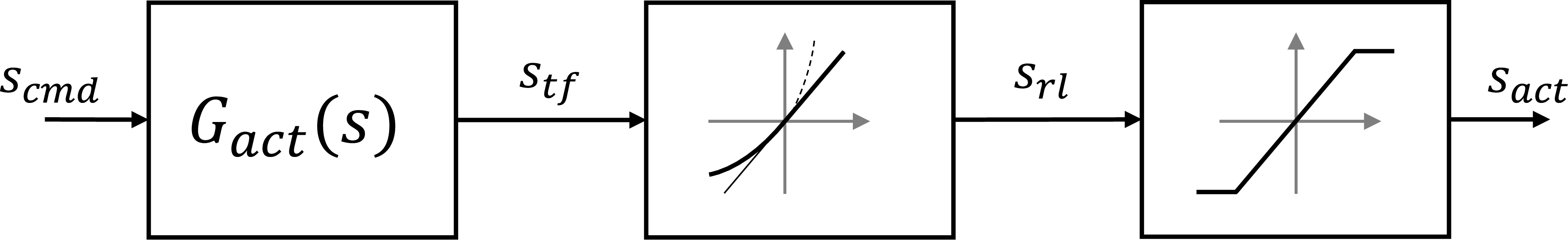}
	\caption{Steer-by-wire actuator model, featuring transfer function, rate limiter and saturation.}
	\label{fig:til_opt_actuator}
\end{figure}
The model considered in this research for the development of benchmark lateral dynamics controllers is a non-linear single-track one (schematically represented in Fig. \ref{fig:til_opt_single_track_model}). Assuming that the right-left tires at front and rear axle can be lumped together, the model reads \citep{beal2012model}
\begin{equation}
	\begin{split}
		\dot{\beta}=&\dfrac{F^f_{y} +F^r_y}{Mv_x}-r\\
		\dot{r}=&\dfrac{L_f F^f_y-L_rF^r_y}{J_{zz}}
	\end{split},
	\label{eq:til_opt_single_track_eqs}
\end{equation}
$J_{zz}$ and $M$ are the vehicle yaw moment of inertia and mass; $L_f$ and $L_r$ are the distance from the vehicle center-of-mass and the front/rear wheels, respectively. The lumped lateral forces at front and rear wheels $F^f_y$, $F^r_y$ can be modeled in different ways; a popular one is the Pacejka magic formula, proposed here in a simplified version \citep{lucchini2020torque}
\begin{equation}
	F_y^{j}=-\dfrac{F_z^{j} \mathcal{C}_j}{\mathcal{A}_j \mathcal{B}_j}\sin \left(\mathcal{B}_j \textrm{atan}\left(\mathcal{A}_j \tan\left(\alpha_j\right)\right)\right),
\end{equation}
whereas $j=f,r$, and $\mathcal{A}_j,\ \mathcal{B}_j,\ \mathcal{C}_j$ are suitable parameters (to be identified from data). $F^j_z$ is the wheel vertical load, $\alpha_j$ is the wheel lateral sideslip angle. Under the assumption of small vehicle sideslip and steer ($s_{act}$) angles, the lateral slip at front and rear tires can be written as
\begin{equation}
	\alpha_f = \beta + \dfrac{L_f}{v_x}r-s_{act},\ \alpha_r = \beta - \dfrac{L_r}{v_x}r.
	\label{eq:til_opt_sideslip_angles}
\end{equation}
On the other hand, vertical forces can be computed by taking into account static contributions (due to vehicle mass), aerodynamic contributions (speed-dependent) and load transfer (acceleration-dependent) \citep{lucchini2021design}, such that
\begin{equation}
	\begin{split}
		F_z^f=&\dfrac{L_r}{L_f+L_r}Mg+k^f_{a}v_x^2-k_xa_x \\
		F_z^r=&\dfrac{L_f}{L_f+L_r}Mg+k^r_{a}v_x^2+k_xa_x
	\end{split}
	\label{eq:til_opt_vertical_forces}
\end{equation}
where $g$ is the gravity acceleration, $k_a^{f},\ k_a^r$ are aerodynamic coefficients, and $k_x$ is the load transfer coefficient. 
\\
\paragraph{Steer-by-wire actuator}
With regards to the steer-by-wire actuator, we consider here a model comprising a linear part (transfer function) and a non-linear one (rate limiter and saturation), see Fig. \ref{fig:til_opt_actuator}. The transfer function model from commanded to actuated steer takes into account the low-level control, and it has been provided by partner company Ferrari S.p.A.
\begin{equation}
	G_{act}\left(s\right)=\dfrac{58.34 s^2 + 1547 s + 9137}{ 1.002 s^3 + 64.55 s^2 + 1549 s + 9137}.
	\label{eq:til_opt_actuator_tf}
\end{equation}
The closed-loop low-level control system modelled by $G_{act}\left(s\right)$ has a bandwidth $\omega_{act}\approx 33.8\ rad/s$. Being $G_{act}\left(s\right)$ complex for the control-oriented modelling purposes herein, we consider a simplified first-order transfer function
\begin{equation}
	G_{act,s}\left(s\right)=\dfrac{\omega_{act}}{s+\omega_{act}}
	\label{eq:til_opt_actuator_simplified}
\end{equation}
The rate-limiter block implements a saturation on the time derivative of the actuated steer angle - it models a current limit in the low-level circuits, specifically. Then, the saturation block implements a limit on the actuated steer angle. Mathematically, this reads
\begin{equation}
	\label{eq:til_opt_actuator_limits}
	\begin{split}
		\dot{s}_{rl} &=
		\begin{cases}
			-\dot{s}_{max} & \text{$\dot{s}_{tf}< -\dot{s}_{max}$}\\
			\dot{s}_{tf} & \text{$-\dot{s}_{max}\leq \dot{s}_{tf}<\dot{s}_{max}$}\\
			\dot{s}_{max} & \text{$\dot{s}_{tf}\geq \dot{s}_{max}$}
		\end{cases}, \\
		s_{act} &=
		\begin{cases}
			-s_{max} & \text{$s_{rl}< -s_{max}$}\\
			s_{act} & \text{-$s_{max}\leq {s}_{rl}<s_{max}$}\\
			s_{max} & \text{$s_{rl} > s_{max}$}
		\end{cases}. \\
	\end{split}
\end{equation}
\paragraph{Complete model and linearization}
The model in Eqs. \eqref{eq:til_opt_single_track_eqs}-\eqref{eq:til_opt_actuator_simplified} contains a non-linearity in the lateral forces $F_y^f$, $F_y^r$. We can linearize their expression for given front and rear sideslip angles $\bar{\alpha}_f,\ \bar{\alpha}_r$
\begin{subequations}
	\begin{align}
		F_y^j \approx &\bar{F}_y^{j}-\bar{C}^j_{\alpha}\left(\alpha_j-\bar{\alpha}_{j}\right), \label{eq:subeq1} \\
		\bar{F}^j_y\left(\bar{\alpha}_j,F_z^j\right)=&-\dfrac{F_z^j \mathcal{C}_j}{\mathcal{A}_j \mathcal{B}_j}\sin \left(\mathcal{B}_j \arctan \left(\mathcal{A}_j \bar{\alpha}_j\right)\right), \label{eq:subeq2} \\
		\bar{C}^j_{\alpha}\left(\bar{\alpha}_j,F_z^j\right)=&\dfrac{F_{z}^j \mathcal{C}_j}{1+\mathcal{A}^2_j \bar{\alpha}_j} \cos \left(\mathcal{B} _j\arctan \left(\mathcal{A}_j \bar{\alpha}_j\right)\right). \label{eq:subeq3}
	\end{align}
\label{eq:til_opt_linearization}
\end{subequations}
Rewriting Eqs. \eqref{eq:til_opt_single_track_eqs}-\eqref{eq:til_opt_actuator_simplified}, and considering Eq. \eqref{eq:til_opt_linearization}, we have the following discrete-time model
\begin{equation}
	\begin{split}
		x_{k+1}=& Ax_{k}+Bu_k+E d_k, \\
		y_k=&Cx_{k}.
	\end{split}
	\label{eq:til_opt_lpv_model}
\end{equation}
Whereas the discretization has been carried out via forward Euler approach, with sampling time $T_s=0.01s$, and the subscript $k$ denotes the time step. The model state $x_k$, output $y_k$, disturbance $d_k$ and input $u_k$ are
\begin{equation}
	\resizebox{0.89 \columnwidth}{!}{$
	\begin{split}
		x_{k}=&\begin{bmatrix}
			\beta_k \\ r_k \\ s_{act,k}
		\end{bmatrix},\ y_k=
		r_k,\ u_k=s_{cmd,k},\\ d_k=&\begin{bmatrix}
			\dfrac{\bar{F}_y^f+\bar{C}^f_{\alpha}\bar{\alpha}_f+\bar{F}_y^r+\bar{C}^r_{\alpha}\bar{\alpha}_r}{Mv_x} \\ \dfrac{L_f\left(\bar{F}_y^f+\bar{C}^f_{\alpha}\bar{\alpha}_f\right)-L_r\left(\bar{F}_y^r+\bar{C}^r_{\alpha}\bar{\alpha}_r\right)}{J_z}
		\end{bmatrix}^t.
	\end{split}
$}
\end{equation}
Whereas $\left(\cdot\right)^{t}$ denotes the matrix transpose operator.
And the model matrices are given in Eq. \eqref{eq:til_opt_model_matrices}.

\begin{figure*}[h]
	\begin{equation}
		\boxed{	\label{eq:til_opt_model_matrices}
			A=T_s\begin{bmatrix}
				T_s^{-1}-\dfrac{\bar{C}^f_{\alpha}+\bar{C}^r_{\alpha}}{Mv_x} & \dfrac{-L_f \bar{C}^f_{\alpha}+L_r \bar{C}^r_{\alpha}-Mv^2_x}{Mv^2_x} & \dfrac{\bar{C}^f_{\alpha}}{Mv_x} \\
				\dfrac{-L_f \bar{C}^f_{\alpha}+L_r \bar{C}^r_{\alpha}}{J_{zz}} & T_s^{-1}-\dfrac{L_f^2 \bar{C}^f_{\alpha}+L_r^2 \bar{C}^r_{\alpha}}{J_{zz}v_x} & \dfrac{\bar{C}^f_{\alpha}L_f}{J_{zz}} \\ 0 & 0 & T_s^{-1}-\omega_{act}
			\end{bmatrix},\ B=\begin{bmatrix}
				0 \\ 0 \\ T_s \omega_{act}
			\end{bmatrix},\ E=\begin{bmatrix}
				T_s & 0 \\ 0 & T_s \\ 0 & 0\end{bmatrix},  C=\begin{bmatrix}
				0 \\ 1 \\ 0\end{bmatrix}^t.}
	\end{equation}
\end{figure*}

\begin{remark}
	\label{remark:til_opt_lpv_model}
	\emph{Note that the model in Eq. \eqref{eq:til_opt_lpv_model} is linear time varying, due to the presence of vehicle speed $v_x$, normal wheel forces $F^f_{z},\ F_z^r$ and the linearization of $F^f_y,\ F^r_y$ around the current sideslip angles. It is common practice to disregard the vehicle speed as a state when dealing with control-oriented modelling for lateral dynamics \citep{beal2012model}, as the same introduces unnecessary complexity; we consider it as a slow-varying parameter. Similar considerations can be applied to the vertical forces \citep{riva_2022_SIL}. Concerning the lateral forces, two possibilities emerge: a first and simpler one consists in linearizing their expression for zero sideslip angles, i.e. $\bar{\alpha}_f=\bar{\alpha}_r=0$ (used e.g. in \citep{corno2023optimization}), the second one consists in linearizing their expression at each time step around the current sideslip angle values (used e.g. in \citep{beal2012model,lucchini2020torque,spielberg_2022}).}
\end{remark}

\subsection{Reference generator}
\begin{figure}[ht]
	\centering
	\includegraphics[width=0.9 \columnwidth]{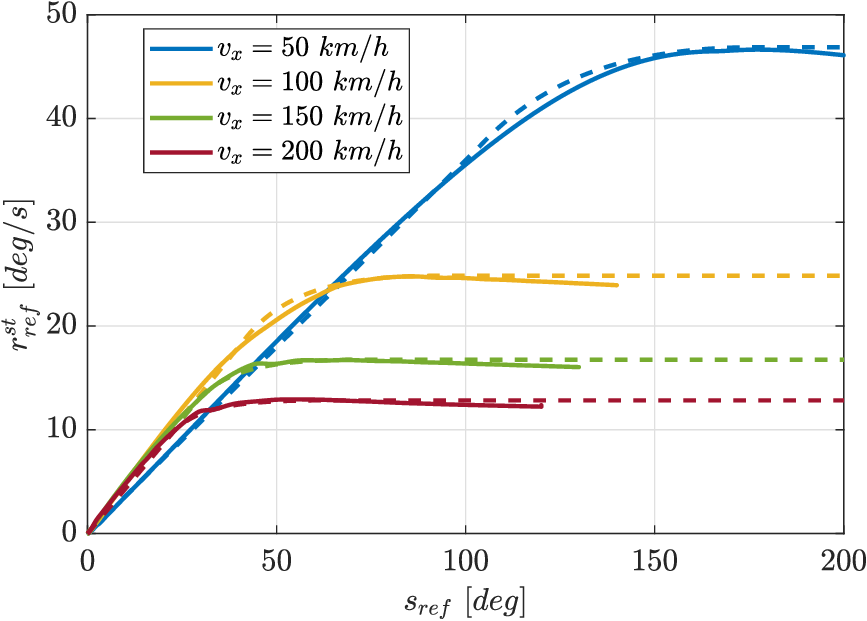}
	\caption{Static component of $r_{ref}$ against driver steer request $s_{ref}$. For each showed speed value, the solid line represents the digital twin behaviour, whereas the dashed one is the static model.}
	\label{fig:til_opt_ref_static}
\end{figure}
\label{Section:til_opt_refgen}
The reference generation approach considered herein transforms a driver steer request into a desired yaw-rate to be imposed onto the vehicle. This can be realized via a static adaptive gain and a transfer function. The static gain defines the desired steady-state vehicle behavior. Considering the available high-fidelity model as a starting point, the reference curves are fitted so as to mimic the nominal vehicle behaviour, saturating the yaw-rate at the maximum obtainable value, see Fig. \ref{fig:til_opt_ref_static}. 
Although the static gain guarantees that the driver request generates a feasible and meaningful yaw-rate reference, the same does not guarantee dynamic feasibility of said reference. Hence, we consider a filter with unitary gain and two poles
\begin{equation}
	F_{ref,dyn}(s)=\dfrac{\left(2\pi f_{ref}\right)^2}{\left(s+2\pi f_{ref}\right)^2}.
\end{equation}
The two poles are set to $f_{ref}=6.3\ Hz$ so as to assign a desired ''bandwidth'' to the controller, as done in \citep{gimondi2021mixed} for a similar problem.

\subsection{Model Predictive Control}
\label{Section:til_opt_mpc}
\begin{figure*}[ht]
	\centering
	\includegraphics[width=0.8 \columnwidth]{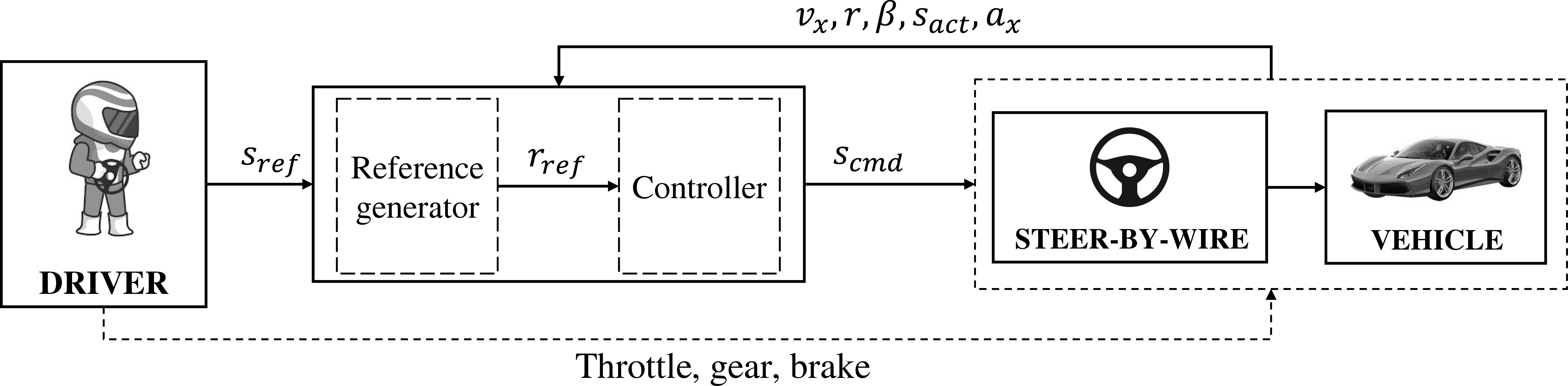}
	\caption{High-level control system architecture, depicting the software elements, namely the reference generator and the controller, the employed measurements and the control variable.}
	\label{fig:til_opt_control_scheme_nominal}
\end{figure*}
\begin{figure}[ht]
	\centering
	\includegraphics[width=1 \columnwidth]{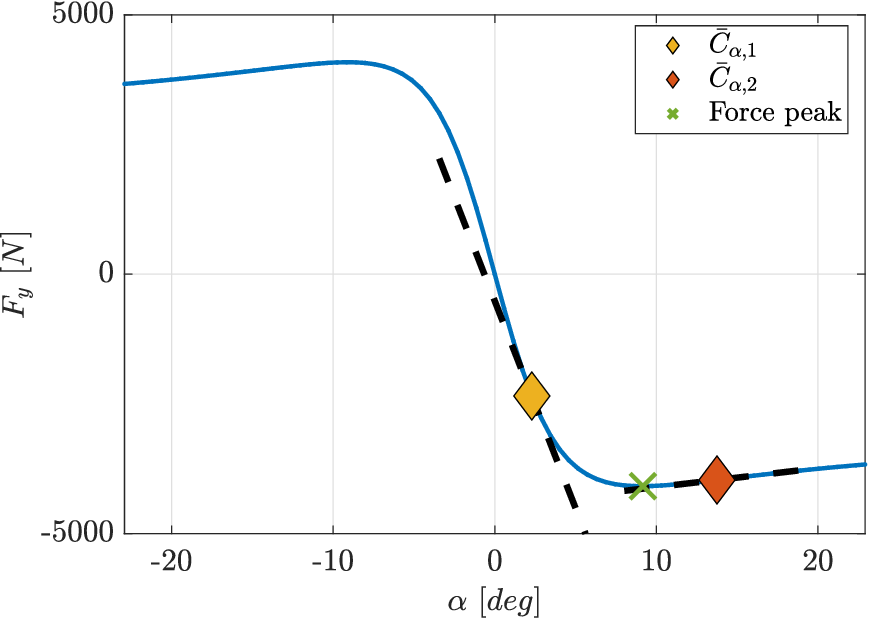}
	\caption{Force-slip diagram with indicated peak force point and two examples of linearized force values before and after the peak.}
	\label{fig:til_opt_peak_force}
\end{figure}
MPC can be employed to optimally solve a constrained control problem. The main ingredients of MPC are an internal model --to be used for predicting the future system behavior at each time step-- a set of constraints on state and input variables, and a cost function to be minimized.
\paragraph{Internal model}
Considering Remark \ref{remark:til_opt_lpv_model}, we apply here the local linearization approach at each time step, for the lateral forces model. Hence, we use as an internal model within the MPC the one described in Eqs. \eqref{eq:til_opt_lpv_model} and \eqref{eq:til_opt_model_matrices}, which is iteratively initialized according to the following procedure
\begin{enumerate}
	\item At time step $k$, measure/estimate $v_{x,k},\ a_{x,k},\  s_{act,k},\ \beta_k$; \\
	\item Use Eq. \eqref{eq:til_opt_sideslip_angles} to estimate $\alpha_{f,k},\ \alpha_{r,k}$. Use Eq. \eqref{eq:til_opt_vertical_forces} to estimate $F_{z,k}^f,\ F_{z,k}^r$;  \\
	\item Linearize lateral forces via Eq. \eqref{eq:til_opt_linearization} to get $\bar{F}^f_{y,k},\ \bar{F}^r_{y,k},\ \bar{C}^f_{\alpha,k},\ \bar{C}^r_{\alpha,k}$;
	\item Substitute Eq. \eqref{eq:til_opt_model_matrices} to get matrix $A_k$.
\end{enumerate}
The model obtained at each time step is thus Linear Time Invariant (LTI), and the MPC requires forward integration of this model to get predictions on the system behavior. For the system to stay LTI, it is necessary that external or potentially time-varying inputs are constant. This means assuming that, for all $k$
\begin{equation}
	\begin{split}
		F^f_{z,k+i}=&F^f_{z,k},\ i=1,\ldots,N_p,\\
		F^r_{z,k+i}=&F^r_{z,k},\ i=1,\ldots,N_p,\\
		v_{x,k+i}=&v_{x,k},\ i=1,\ldots,N_p.
	\end{split}
\end{equation}
The assumptions above are indeed common in the literature \citep{lucchini2020torque,beal2012model}. A special note is necessary here for what concerns the lateral force model. Consider the force-slip relation for the considered front tires, in Fig. \ref{fig:til_opt_peak_force}: it is evident that the maximum attainable force is achieved for $\alpha=\alpha_{max}$, much smaller than $ s_{max}$. Going above the force peak, the slope of the curve is inverted -- \emph{i.e.} $\bar{C}_{\alpha}$ changes sign -- and the predictive controller would try to increase $\alpha$ to reduce the lateral force, given its limited knowledge of the force model; one can verify that this yields instability. While accounting for the complete force model is non-viable for the introduced nonlinearities, enforcing a constraint on $\alpha$ allows for smooth operation of the tire within its physical limits.
Although a human driver might want to stay beyond the curve for fun-to-drive reasons -- \emph{e.g.} for increased slipping of the vehicle -- no advantages in terms of steering performance exist. Let us remark that this consideration was not carried out in previous literature \citep{lucchini2020torque,lucchini2021design}, but is non-viable when dealing with driving at the limits of handling. Since we are considering active front steering, only the front sideslip angle $\alpha_f$ is constrained.

Indeed, constraining $\alpha$ requires good knowledge of the slip value for which force generation is maximized: imperfect knowledge might yield to worse control performance, and an ad-hoc learning approach to estimate this parameter in real-time should be used (\emph{e.g.} a neural-network, as in \citep{spielberg_2022}). For the present research, $\alpha_{max}$ is assumed to be known with adequate accuracy.
\paragraph{Constraints}
The system in Eq. \eqref{eq:til_opt_lpv_model} is characterized by physical limitations on the actuator (Eq. \eqref{eq:til_opt_actuator_limits}). Said limitations are to be included as hard constraints within the optimization problem.
\begin{equation}
			\resizebox{0.99 \columnwidth}{!}{$
	\label{eq:til_opt_actuator_constraints_mpc}
	\begin{split}
		&-s_{max}\leq s_{act,k} \leq s_{max},\ \forall k=1,\ldots,N_p\\
		&-\dot{s}_{max} T_s \leq s_{act,k}-s_{act,k-1} \leq \dot{s}_{max} T_s,\ \forall k=1,\ldots,N_p.
	\end{split}
			$}
\end{equation}
On the other hand, as discussed above, the front sideslip angle $\alpha_f$ should be limited to prevent the system from going above the tire limitations. This not being a physical limit of the system (which is always able to enter the above-peak force region), we implement it as a soft constraint, via slack variable $\rho$
\begin{equation}
	\label{eq:til_opt_soft_constraint}
	\begin{split}
		&-\alpha_{f,max}-\rho\leq \alpha_{f,k}\leq \alpha_{f,max}+\rho,\ \forall k=1,\ldots, N_p
	\end{split}
\end{equation}
As discussed above, the constraint based on maximum $\alpha_{f,k}$ is much more restrictive than the actuator capabilities (constraints in Eq. \eqref{eq:til_opt_actuator_constraints_mpc}). Hence, first constraint in Eq. \eqref{eq:til_opt_soft_constraint} can be discarded, as it is never activated.
\paragraph{Cost function}
\begin{figure}[th]
	\begin{equation}
		\resizebox{0.99 \columnwidth}{!}{$
			\boxed{\label{eq:til_opt_opt_problem_MPC} 
				\begin{split}
					\mathcal{U}*= &\argminF_{\epsilon, {x}_1,\ldots,{x}_{N_p}, u_0,\ldots,  u_{N_p}}  \sum_{k=0}^{N_p-1} { {x}^t_{k+1}}W_x { {x}_{k+1}}+w_u u_k^2+w_{\rho} \rho^2 ,\\
					\textrm{subject to} & \\
					x_0 =& \begin{bmatrix} \beta_0 &  r_0 &  s_0 \end{bmatrix}^t,\\			
					{x}_{k+1} =& A {x}_k + B  u_k + E  d_k,\ \forall k=0,\ldots,N_p-1 \\	
					-\dot{s}_{max} T_s \leq& \delta s_k \leq \dot{s}_{max} T_s,\ \forall k=1,\ldots,N_p \\
					-\alpha_{f,max}-\epsilon\leq& \alpha_{f,k} \leq \alpha_{f,max}+\epsilon,\ \forall k=1,\ldots,N_p.
				\end{split}
			}
			$}
	\end{equation}
\end{figure}
The minimized cost is given in Eq. \eqref{eq:til_opt_opt_problem_MPC}, together with the constraints. More specifically, the state vector is weighted ($W_x$), while a penalty on the control effort ($w_u$) avoids numerical instabilities, and the slack variable is minimized ($w_{\rho}$) to ensure that $\rho=0$ when inside the constraint limits. 
Given that the prediction model is locally Linear Time-Invariant, the Single Shooting Approach can be applied to write the states, the constraints and the cost metric as a function of the control inputs (\emph{e.g.} as in \citep{riva2022model}). The obtained optimization problem is a Quadratic Programming one, fastly solved via benchmark tools -- \textit{qpOASES} is used in this research \citep{Ferreau2014}.

\section{Twin-in-the-Loop Control}
\label{Section:til_opt_til_control}
\begin{figure*}[ht]
	\centering
	\includegraphics[width=0.8 \columnwidth]{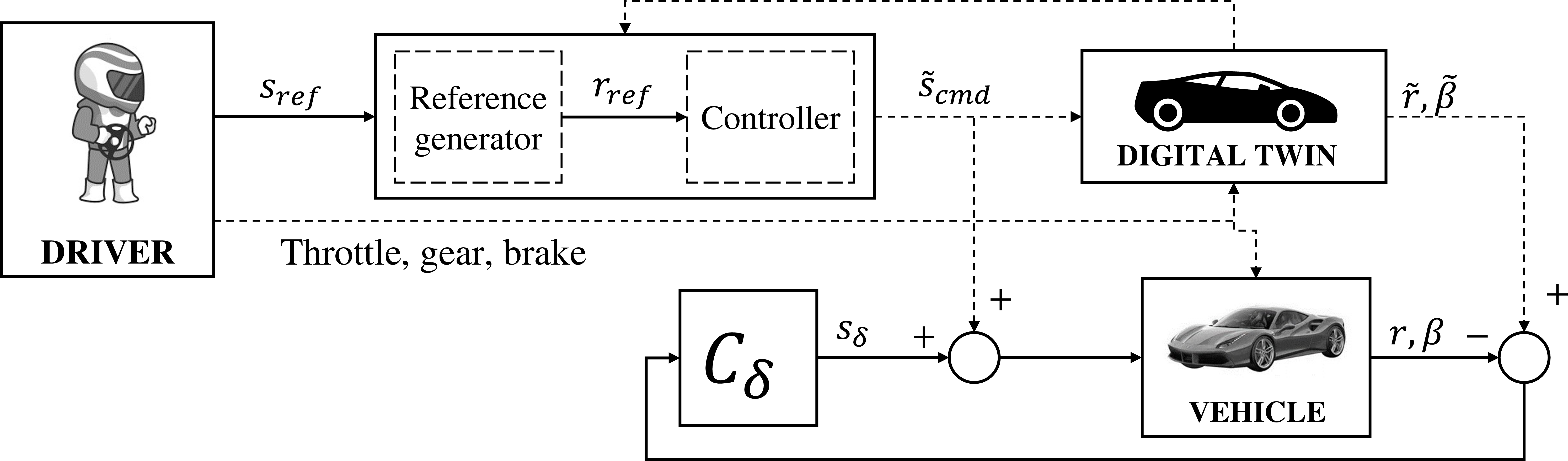}
	\caption{TiL control scheme for yaw-rate tracking.}
	\label{fig:til_opt_silc_scheme}
\end{figure*}
The TiL scheme of Figure \ref{fig:til_opt_silc_scheme} is considered in this Section. The architecture features the MPC controller of Section \ref{Section:til_opt_mpc}, a DT and a compensator $C_{\delta}$. 

As stated in \citep{dettu2023twin}, under the hypothesis that the simulator is a faithful replica of the real system, $C_{\delta}$ controls a small signal dynamic, as the open-loop steering action from the DT $\tilde{s}_{cmd}$ does most of the job. Hence, we consider a simple PID compensator.
Leveraging findings in \citep{gimondi2021mixed}, we employ here a mixed yaw-rate sideslip control, whereas the regulator $C_{\delta}$ controls a mixed signal, rather than the yaw-rate per-se
\begin{equation}
	\epsilon = (1-\zeta)r-\zeta \beta,\ \zeta\in \left[0,1\right]
	\label{eq:til_opt_c_delta}
\end{equation}
Introducing a correction onto the yaw-rate via $\beta$ yields better results in case of very high cornering stiffness values, and allows to account for $\beta$ within the Single Input Single Output controller. The control law, in continuous time ($s$ denotes the Laplace domain variable), reads
\begin{equation}
	C_{\delta}\left(s,\theta \right)=k_p\left(1+\dfrac{1}{sT_I}+\dfrac{sT_D}{1+sT_D/N_D}\right).
\end{equation}
$\theta=\left\{k_p, T_I, T_D\right\}$ is the vector of tuning knobs, and the high-frequency pole is set ten times faster ($N_D=10$) than $T_D$, as in common practice. The controller is discretized via the Tustin approach, with sampling time $T_s=0.01s$.

Following the peak force limit considerations in Section \ref{Section:til_opt_mpc}, we enforce a saturation on the actuated steer angle
\begin{equation}
	s^{lb}_{max,d}- \tilde{s}_{cmd}\leq s_{\delta}\leq s^{ub}_{max,d} - \tilde{s}_{cmd}.
\end{equation}
Whereas $s^{lb}_{max,d},s^{ub}_{max,d}$ are obtained by substituting in Eq. \eqref{eq:til_opt_sideslip_angles} the maximum sideslip angle $\alpha_{f,max}$, used in the MPC constraints.
\subsection{$C_{\delta}$ tuning}
\label{Section:til_opt_c_delta_tuning}
To calibrate $C_{\delta}$, we proposed an optimization problem, solved via Bayesian Optimization (BO) in \citep{dettu2023twin}: this is necessary given that the block controls the unknown residual dynamic between the vehicle and its replica. However, performing experiments with black-box proposed parameters might lead to instability or unsafe configurations; we introduce here a constraint on $\beta$, to be satisfied during the optimization process. This guarantees that while exploring the search space, the algorithm stays away from high sideslip values --- note that using the sideslip as an indicator of vehicle maneuverability and safety is common in the literature \citep{gimondi2021mixed,beal2012model}.
\begin{equation}
	\label{eq:til_opt_optimization_problem}
	\begin{split}
		\theta^* =& \argmin_{\theta}f_{bo}\left(\theta\right) \\
		\textrm{s.t.}\ \ \ & \theta \subseteq \Theta \in \mathbb{R}^{N_{\theta}} \\
		&  g_c\left(\theta\right) \geq 0
	\end{split}
\end{equation}
Where $g_c\left(\theta\right)=-\left|\beta\left(\theta\right)\right|+\beta_{max}$ is the constraint.
The cost function $f_{bo}$ reads
\begin{equation}
	f_{bo}(\theta)= \dfrac{1}{N_{exp}}\sum_{k=1}^{N_{exp}} \left(\tilde{\epsilon}_k-\epsilon_k\left(\theta\right)\right)^2+\gamma_u \left(\dot{{s}}_{cmd,k}\right)^2
	\label{eq:til_opt_cost_BO}
\end{equation} 
Where the first term on the right hand side weights the controller tracking performance, and the second one weights the control action derivative, similarly to what done in Eq. \eqref{eq:til_opt_opt_problem_MPC} for the MPC. $N_{exp}$ is the number of samples in the considered experiment.
As for the cost function, also the constraint on $\beta$ is a black-box, and needs to be adequately treated. Constrained BO \citep{khosravi2022safety} can been used to solve the problem. CBO estimates the black-box constraint via a Gaussian Process, and the availability of a set of data. Analogously to cost function regression, the optimizer iteratively adjusts its knowledge of the constraint.
In practice, at each $n$-th iteration, the black-box optimizer evaluates $\theta^{\left(n\right)}$, and the tuple $\left(\theta^{\left(n\right)},f_{bo}^{\left(n\right)},g^{\left(n\right)}_c\right)$ is collected and used to increment the dataset $\mathcal{X}^{n}=\mathcal{X}^{n-1} \cup \left(\theta^{\left(n\right)},f_{bo}^{\left(n\right)},g^{\left(n\right)}_c\right)$. The dataset contains data about the unknown cost function and constraint, allowing estimation of the same.\\
Although BO proved to be effective in solving the problem, it still requires many iterations to converge to an optimal solution; given that performing an experiment with the vehicle might be costly, we would like to assess whether a faster convergence can be achieved by selecting different optimization tools. Also, the high computational burden of BO is one important issue, and has been recently tackled \citep{sabug2022smgo}.
\subsubsection{SMGO-$\Delta$}
Set-Membership Global Optimization-$\Delta$ (SMGO-$\Delta$) is a novel data-driven optimization, featuring unknown constraints \citep{sabug2022smgo}. Given a set of samples of the objective and constraints, SMGO-$\Delta$ uses the existing information to build a surrogate Set Membership-based model of the hidden functions, and iteratively selects the next point to evaluate. Such a model is used to strategically decide between performing an exploitation sampling in the vicinity of the current best
sample, or exploration sampling around the search space to discover the function shape. SMGO-$\Delta$ can be used to solve a problem of the type in Eq. \eqref{eq:til_opt_optimization_problem}. 
We give here some details of the algorithm functioning; the reader is referred to \citep{sabug2022smgo} for detailed information.

By iteratively collecting data about the cost function and constraints as detailed above, SMGO calculates Lipschitz constant $\gamma_{f,g_c}^{\left(n\right)}$ and noise bound $\varepsilon_{f,g_c}^{\left(n\right)}$ estimates for both $f_{bo}$ and $g_c$ are computed and then used to estimate upper and lower confidence bounds --- note that BO attempts at fitting a surrogate of $f_{bo}$ and $g_c$, while SMGO mostly considers the confidence bounds, thus being much lighter. As an example, for $f_{bo}$, we have that
\begin{equation}
	\begin{split}
		\obar{f}_{bo}^{\left(n\right)}\left(\theta\right)=&\min_{k=1,\ldots,n}\left(f_{bo}^{\left(k\right)}+\gamma^{\left(n\right)}\left|\left|\theta- \theta^{\left(k\right)}\right| \right| \right),\\
		\ubar{f}_{bo}^{\left(n\right)}\left(\theta\right)=&\max_{k=1,\ldots,n}\left(f_{bo}^{\left(k\right)}-\gamma^{\left(n\right)}\left|\left|\theta- \theta^{\left(k\right)}\right| \right| \right).
	\end{split}
\end{equation}
The upper and lower bounds are then used to compute the mean value $\tilde{f}_{bo}^{\left(n\right)}$ and the associated uncertainty $\lambda^{(n)}\left(\theta\right)$
\begin{equation}
	\begin{split}
		\lambda^{(n)}\left(\theta\right)=&\obar{f}_{bo}^{(n)}\left(\theta\right)-\ubar{f}_{bo}^{(n)}\left(\theta\right),\\
		\tilde{f}_{bo}^{(n)}\left(\theta\right)=&\dfrac{1}{2}\left(\obar{f}_{bo}^{(n)}\left(\theta\right)+\ubar{f}_{bo}^{(n)}\left(\theta\right)\right)
	\end{split}
\end{equation}
The algorithm then selects the next point to evaluate $\tilde{\theta}^{(n)}$ based on the following problem
\begin{equation}
	\begin{split}
		\tilde{\theta}^{(n)}=&\argmin_{\theta \in \mathcal{S}^{(n)}} \tilde{f}_{bo}^{(n)}\left(\theta\right)-\beta \lambda^{(n)}\left(\theta\right) \\
		\textrm{s.t.} \ \ & \Delta \tilde{g}^{(n)}\left(\theta\right) +\left(1-\delta\right) \ubar{g}^{(n)}\left(\theta\right)\geq 0
	\end{split}
\end{equation}
Whereas $\beta$ trades off between \textit{exploitation} ($\beta=0$) and \textit{exploration} ($\beta \geq 0$), and $\Delta$ trades off between \textit{cautiousness} ($\delta=0$) and \textit{riskiness} ($\delta=1$) in constraint exploration. The candidate point $\tilde{\theta}^{(n)}$ is then used to assess the inequality
\begin{equation}
	\ubar{f}_{bo}^{(n)}\left(\tilde{\theta}^{(n)}\right) \leq f_{bo}^{*\left(n\right)}-\alpha_{smgo} \gamma^{(n)}_f
	\label{eq:til_opt_smgo_test}
\end{equation}
If Eq. \eqref{eq:til_opt_smgo_test} above is satisfied, the algorithm effectively employs $\tilde{\theta}^{(n)}$ as the next point to evaluate, otherwise, the \textit{exploration} routine is triggered, which aims at reducing the uncertainty (the right hand side of Eq. \eqref{eq:til_opt_smgo_test}) while satisfying the constraints.

\subsubsection{Virtual Reference Feedback Tuning}
\begin{figure}[ht]
	\centering
	\includegraphics[width=1 \columnwidth]{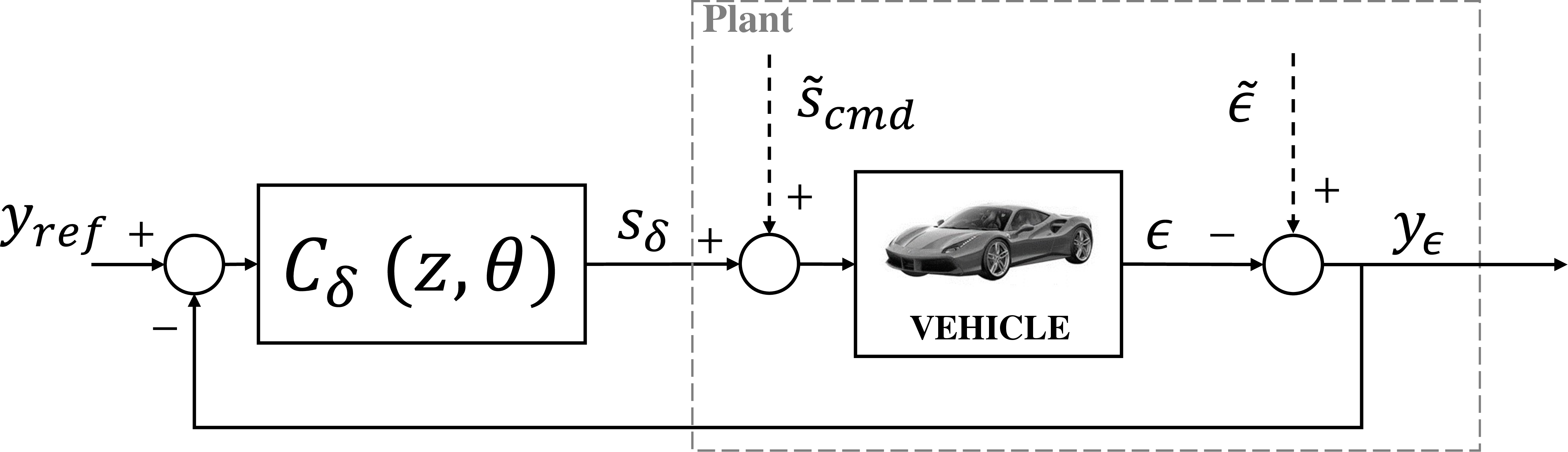}
	\caption{Reworked TiL control scheme, for VRFT implementation.}
	\label{fig:til_opt_vrft_scheme}
\end{figure}
VRFT is a direct data-driven method for the design of Single-Input Single-Output (SISO) parametric controllers, that does not require plant knowledge \citep{campi2002virtual}. It selects a controller, parametrized by $\theta$, by minimizing the following cost (whereas the controller is linear in this setup)
\begin{equation}
	\resizebox{0.99 \columnwidth}{!}{$
		\theta^*_{vrft}\left(\theta\right)=\argmin_{\theta}\left|\left|  \left( \dfrac{P\left(z\right)C_{\delta}\left(\theta\right)}{1+P\left(z\right) C_{\delta} \left(\theta\right)}-M_r\left(z\right) \right)M_w\left(z\right) \right| \right|_2^2.
		\label{eq:til_opt_vrft_cost}
		$}
\end{equation}
The cost in Eq. \eqref{eq:til_opt_vrft_cost} means that VRFT tries to find a controller matching in the frequency domain a discrete-time reference model $M_r(z)$, given a weighting function $M_w(z)$.
Indeed, theoretical guarantees on Eq. \eqref{eq:til_opt_vrft_cost} only apply to linear plants: in our specific case, no linearity assumption can be drawn based on the plant in Fig. \ref{fig:til_opt_vrft_scheme} (the plant encompasses both the digital twin and its interaction with the controller). However, as showed in \citep{dettu2023twin}, if the digital twin is a good replica of the real system, we can reasonably assume that the dynamics to be controlled is a small signal one. Nonetheless, VRFT has been succesfully used in many real systems, where the ''linearity'' concept looses its meaning \citep{busetto2023data,radrizzani2020data,passenbrunner2014direct}.

Being the plant unknown, Eq. \eqref{eq:til_opt_vrft_cost} is not usable; only sequences of input $\left(s_{\delta,1},\ldots,s_{\delta,N_{exp}}\right)$ and output $\left(y_{\epsilon,1},\ldots,y_{\epsilon,N_{exp}}\right)$ data are available (an open-loop experiment suffices). Hence, the cost function in Eq. \eqref{eq:til_opt_vrft_cost} is suitably rewritten as a function of available data only \citep{care2019toolbox}
\begin{equation}
	\resizebox{0.99 \columnwidth}{!}{$
		f_{vrft}\left(\theta\right)=\dfrac{1}{N_{exp}}\sum_{k=1}^{N_{exp}}M_w(z)\left(C_{\delta}\left(\theta\right)\left(M_r(z)^{-1}-1\right)y_{\epsilon,k}-s_{\delta,k}\right).
		\label{eq:til_opt_vrft_real_cost}
		$}
\end{equation}
In the TiL setup, regulator $C_{\delta}$ controls the error $y_{\epsilon}=\tilde{\epsilon}-\epsilon$, and regulates it to zero. 
\subsubsection{Comparing VRFT with global optimization}
%
The obtained control parameters are then prone to be used as an initial guess for a successive iterative algorithm (\emph{e.g.} as in \citep{busetto2023data}).  
As one could note, the cost function of VRFT Eq. \eqref{eq:til_opt_vrft_real_cost} is markedly different from the one showed in Eq. \eqref{eq:til_opt_optimization_problem}. VRFT requires a reference model $M_r(z)$, and cannot arbitrarily learn a controller minimizing the tracking error, as for global optimizers. Apart from the reference model, to enhance performance is often necessary to consider a $M_w(z)$, reasonably setting the frequencies of interest. Nonetheless, we will show further on as the solution found by VRFT is well performing if evaluated onto the cost of Eq. \eqref{eq:til_opt_cost_BO}.

Also, global optimizers allow for an arbitrary cost function, \emph{e.g.} taking into account the control effort (see Eq. \eqref{eq:til_opt_optimization_problem}): this is not possible in classic VRFT.
For a fair comparison of VRFT with BO, one could execute the optimization routine described in Section \ref{Section:til_opt_c_delta_tuning} by using the VRFT cost in Eq. \eqref{eq:til_opt_vrft_real_cost}, which in practice means pre-filtering the data according to $M_r(z)$ and $M_w(z)$, and removing the weight onto $\dot{s}_{act}$. VRFT is here applied onto the tracking error $y_{\epsilon}$, and requires a virtual reference to be passed through the desired model $M_r(z)$. However, the reference for the error should be zero in a closed-loop setup like the one considered in BO and SMGO, for this reason, the global optimizers cost function is somewhat different in nature, in that it minimizes the $rms$ of the error in a specific closed-loop experiment, with zero reference.
\begin{equation}
	f_{bo-vrft}(\theta)= \dfrac{1}{N_{exp}}\sum_{k=1}^{N_{exp}} M_w(z)M_r(z)y_{\epsilon,k}^2
	\label{eq:til_opt_cost_BO_vrft_like}
\end{equation} 
\section{Simulation results}
\label{Section:til_opt_simulation_results}
In the following we conduct a series of simulation tests to assess the performance of TiL lateral control onto the vehicle, while also analysing and comparing the different optimization tools described above.
\subsection{Simulation setup}
\subsubsection{The vehicle and the digital twin}
\begin{table}
	\centering
	\begin{adjustbox}{width=\columnwidth}
		\begin{tabular}{@{}lllll@{}}
			\toprule
			$M\ [kg]$           & $J_{zz}\ [kg m^2]$      & $L_{f,r}\ [m]$         & $\mathcal{A}_{f,r}$      & $\mathcal{B}_{f,r}$ \\ \midrule
			$1729.1$            & $2482.7$                & $1.48,\ 1.16$    & $10.72,\ 19.75$          & $1.51,\ 0.75$       \\ \midrule
			$\mathcal{C}_{f,r}$ & $k_a^{f,r}\ [kg/m]$ & $k_x\ [kg]$ & $\dot{s}_{max}\ [deg/s]$ & $\alpha^f_{max}\ [deg]$    \\ \midrule
			$20.08,\ 28.69$            &   $0.065,\ 0.221$      & $153.63$      & $100$          & $9.10$       \\ \bottomrule
		\end{tabular}
	\end{adjustbox}
	\caption{Set of vehicle parameters for the simplified model of Section \ref{Section:til_opt_control_oriented_model}.}
	\label{tab:til_opt_vehicle_parameters}
\end{table}
For testing the TiL controller, since we do not have the possibility of performing experiments on an actual vehicle, a different instance of the Digital Twin is generated and regarded as the "Physical Vehicle".
The considered Digital Twin is modeled in \textit{VI-CarRealTime} (an off-the-shelf commercial simulation software \citep{vigrade_2022}). It models an high-performance two seats car, and the most relevant parameters for the problem under analysis are given in Table \ref{tab:til_opt_vehicle_parameters}; they have been provided by partner company Ferrari S.p.A. Note that the displayed parameters are those of the simplified model derived in Section \ref{Section:til_opt_control_oriented_model} and used within the MPC model. The true underlying model features much more parameters; \emph{e.g.} version 5 of the Magic Formula is used for modelling the tire-road interaction, while we used a 3-parameters approximation of it within the MPC, encompassing lateral forces only.

We assume that the real vehicle features some different elements and non-idealities, which are absent in the Digital Twin:
\begin{itemize}
	\item The vehicle has real sensors, characterized by noise. Zero-mean Gaussian distributed noise is added onto the yaw-rate measure $r$, characterized by standard deviation $\sigma_{n,r}=0.006\ rad/s$. Similarly for the sideslip, we mimic the presence of state-estimator, introducing a low-pass filtered noise with standard deviation $\sigma_{n,\beta}=0.0044\ rad$. Overall, this yields an $snr$ of $\approx 20$ in the considered signals, and the comparison between real and noisy $\beta$ is similar to what obtained in recent works on the topic \citep{carnier2023hybrid};
	\item Additional masses are modelled in the real vehicle. Specifically, we add a $100\ kg$ passenger and an unbalanced load on the front trunk ($70\ kg$ on the left, $10\ kg$ on the right). The layout is the same considered in \citep{dettu2023twin}; 
	\item We reduce the rear wheels cornering stiffness by the $15\ \%$, so as to mimic a wrong tire model within the model based controller, or a modification of the tire property due \emph{e.g.} to low inflation pressure.
\end{itemize}
The modifications above reproduce a realistic case study, on which the optimization is run and the controllers validated.

\textit{VI-CarRealTime} is provided with a virtual driver feature, which is here used to maintain a given speed profile (working as a longitudinal dynamics controller): indeed, depending on the vehicle lateral dynamics, slightly different speed profile tracking might be achieved
\subsubsection{Parameters selection}
We here briefly describe the parameters selection for the various controllers and optimization algorithms used. Concerning the MPC, the following weights are chosen; they have been found through fine tuning onto the Digital Twin 
\begin{equation}
	w_u = 1,\ w_{\rho} = 100,\ w_{\beta} = 0.2,\ w_r=0.8,\ w_s = 0.
\end{equation}
The same are used within the TiL control scheme of Fig. \ref{fig:til_opt_silc_scheme}. In $C_{\delta}$, the parameter $\zeta$ (see Eq. \eqref{eq:til_opt_c_delta}) trades off yaw-rate tracking for sideslip minimization. It is set to $0.2$ so as to focus on the first objective, while avoiding unsafe sideslip increments.

With regards to SMGO, $\Delta=0.5$ is used, to guarantee a balanced trade-off between cautiousness and riskiness in constraint exploring. Other tuning parameters for SMGO are let at their default values \citep{sabug2022smgo}.

With regards to CBO, default values, the exploration ratio is set to $0.5$, and the acquisition function is set as the Expected Improvement one.
$\beta_{max}$ is set at $4.5\ deg$, which is $\approx 2.5\ deg$ more than what achieved by using the nominal MPC in the optimization maneuver: we want not to significantly exceed this limit while training the controllers onto the same maneuver. Indeed, this is not an issue for different maneuvers, where the sideslip might naturally exceed this limit also in nominal conditions, and $\beta_{max}$ has to be chosen for the specific maneuver of interest.

VRFT requires tuning of reference model $M_r(z)$ and frequency weighting function $M_w(z)$. The first one is taken from the literature \citep{gimondi2021mixed}, where a controller for yaw-rate dynamics is closed at the bandwidth of $3.5\ Hz$: we thus set it as a first-order low-pass filter, with one pole at this frequency. The frequency range is instead taken from the reference generation filter (see Section \ref{Section:til_opt_refgen}); since the reference is cut after $f_{ref}=6.3\ Hz$, we assign the same dynamics to $M_w(z)$. Of course, both reference model and weight function are suitably discretized via the Tustin approach.
\subsection{Optimization results comparison}
\begin{figure}[ht]
	\centering
	\includegraphics[width=1 \columnwidth]{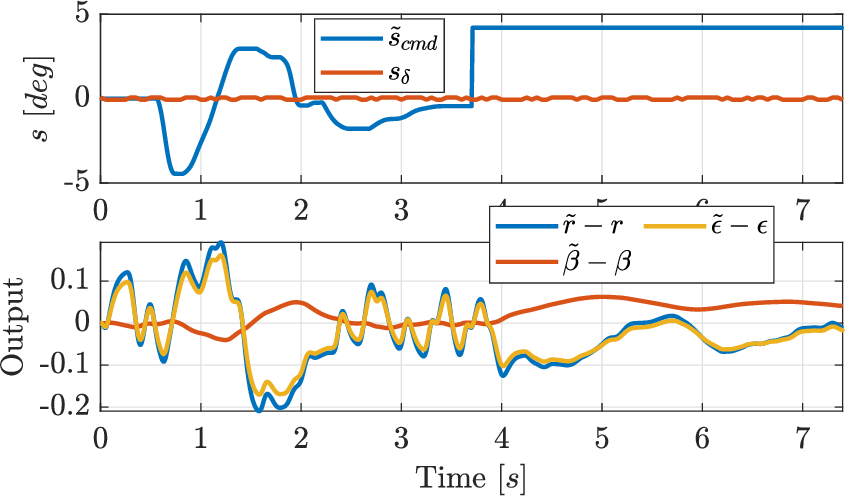}
	\caption{Input and output signals for VRFT use.}
	\label{fig:til_opt_vrft_test}
\end{figure}
\begin{figure}[ht]
	\centering
	\includegraphics[width=1 \columnwidth]{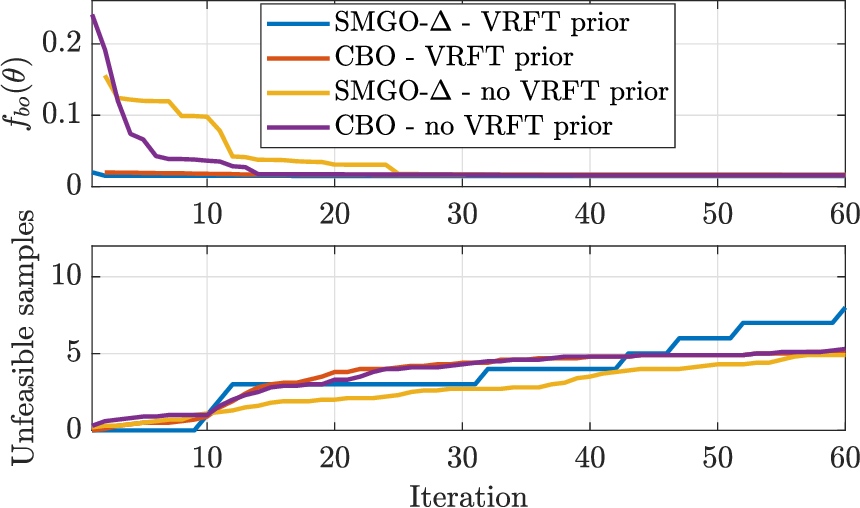}
	\caption{Comparison of black-box optimization algorithms for TiL calibration.}
	\label{fig:til_opt_calibration}
\end{figure}
\begin{figure}[ht]
	\centering
	\includegraphics[width=1 \columnwidth]{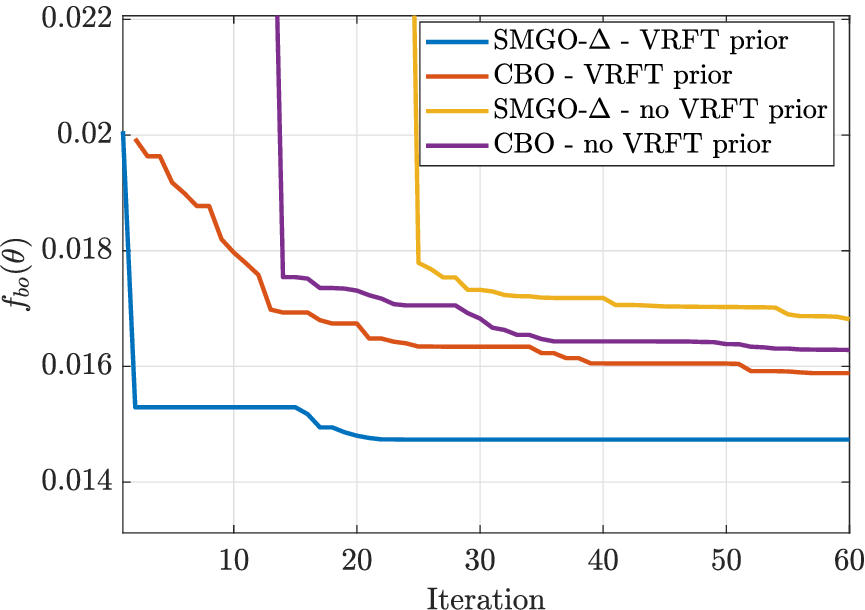}
	\caption{Comparison of black-box optimization algorithms for TiL calibration. Highlight on the cost function.}
	\label{fig:til_opt_calibration_highlight}
\end{figure}

\begin{table}
	\resizebox{\columnwidth}{!}{
		\begin{tabular}{llll}
			\hline
			\textbf{Controller} & $rms(\tilde{r}-r)$ $\left[deg/s\right]$ & $rms(\tilde{\beta}-\beta)$ $\left[deg\right]$ & $rms(\dot{s}_{act})$ $\left[deg/s\right]$ \\ \hline
			MPC                 & $1.85$                                 & $1.81$                                     & $918.79$                               \\ \hline
			TiL-SMGO-$\Delta$               & $0.74$                                 & $0.86$                                     & $514.41$                               \\ \hline
			TiL-VRFT            & $1.06$                                 & $0.95$                                     & $475.47$                               \\ \hline
			TiL-SMGO-$\Delta$  (VRFT cost)  & $1.04$                                 & $0.89$                                     & $935.42$                               \\ \hline
		\end{tabular}
	}
	\caption{Performance indices of the controllers evaluated in the optimization test (double-lane-change with step-steer at $120\ km/h$).}
	\label{tab:til_opt_optimization_test_metrics}
\end{table}

For the optimization of $C_{\delta}$, we consider a double-lane change maneuver, followed by a step steer, executed at $120\ km/h$. \\
To use VRFT, we execute the maneuver onto the digital twin (\emph{i.e.} on the upper branch in Fig. \ref{fig:til_opt_silc_scheme}), while feeding the physical vehicle with a Pseudo Random Binary Input signal (as in \cite{radrizzani2020data}), to excite the dynamics to be controlled. These information is displayed in Fig. \ref{fig:til_opt_vrft_test}; the upper plot depicts the command from DT $\tilde{s}_{cmd}$ and the one fed in open-loop $s_{\delta}$. The lower plot depicts the system outputs $r$ and $\beta$, as well as their combination $\epsilon$.

The learned controller is then used as a prior for global optimization; each optimization iterates for $60$ steps. To minimize the effect of randomization present in both CBO and SMGO-$\Delta$, we perform each optimization $10$ times. The results are displayed in Fig. \ref{fig:til_opt_calibration}: in the upper plot, the optimal cost found at each iteration (averaged per the $10$ optimization runs) is displayed.

The lower plot instead displays the number of infeasible experiments for each controller, at each iteration, \emph{i.e.} breaking the constraint on $\beta$. 
From the upper plot, we note as VRFT is capable of delivering an almost optimal solution, with just one experiment; BO and SMGO are able to approach VRFT cost after $>10$ iterations on average. Figure \ref{fig:til_opt_calibration_highlight} shows an highlight of the upper plot: SMGO provides the better results in terms of cost function minimization, while CBO is somewhat less performing. In general, both optimizers fail at converging at the best point when VRFT prior is not provided. It is interesting to note that these results confirm the findings in \citep{busetto2023data}, for a different application.\\
If one looks at the bottom plot is clear as the increased performance achieved with SMGO is traded with somewhat less cautious exploration, in that the number of unfeasible attempts grows. This is consistent to what found in \citep{busetto2023data}; when provided with VRFT prior, SMGO fastly switches from exploitation mode to exploration mode (as the global minimum is in practice already achieved by using VRFT), and this eventually translates into breaking the constraints more easily.

Concerning the computation time, at each iteration CBO takes on average \footnote{These results have been obtained with MATLAB 2022a, on a $16GB$ RAM Asus Laptop, with an Intel Core i7-8750H $2.20GHz$ processor.} $0.4442\ s$ --with $0.0961\ s$ of standard deviation-- to compute the next point to be evaluated; SMGO-$\Delta$ is on the other hand significantly faster, taking $0.0238\ s$ on average, with $0.0017\ s$ of standard deviation. In practice, SMGO-$\Delta$ is almost two orders of magnitude faster than CBO for this optimization problem -- consistently to what found in \citep{sabug2022smgo}, and this potentially allows for the whole tuning procedure to be run on low cost computing units and at real-time.
\subsection{Time-domain results comparison}
In the following, we compare the controllers optimized as above in the time-domain. Being CBO and SMGO-$\Delta$ calibrations very similar, we show here the second one. Also, given that the two global optimizers are able to converge only by using VRFT control parameters as a prior, we consider the corresponding SMGO-$\Delta$ calibration initialized with VRFT.
We then show the results obtained via VRFT calibration, and for the optimization maneuver, we also show what would happen if explicitly using the VRFT cost function within SMGO-$\Delta$.
\subsubsection{Double lane change with step steer - $120\ km/h$}
\begin{figure}[ht]
	\centering
	\includegraphics[width=1 \columnwidth]{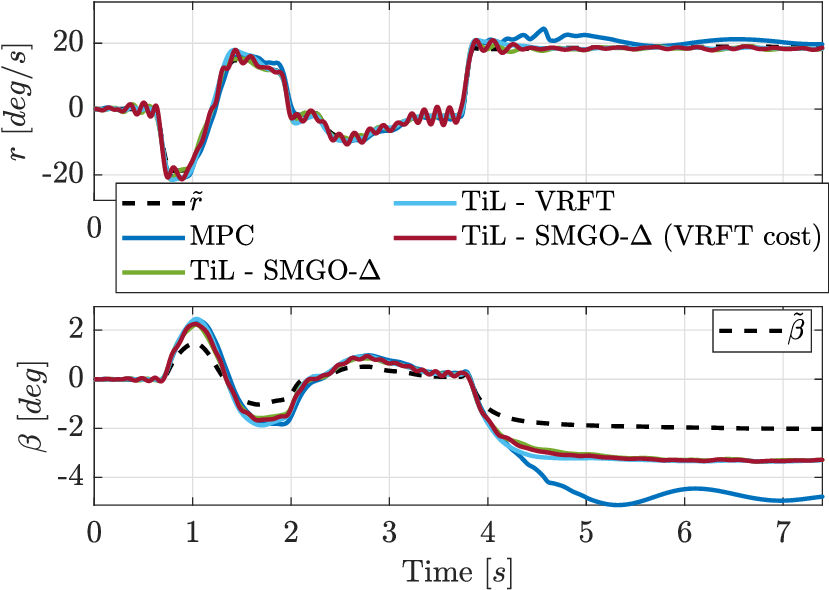}
	\caption{Controller validation onto the optimization experiment: yaw-rate and sideslip tracking.}
	\label{fig:til_opt_optimization_experiment_tracking}
\end{figure}

\begin{figure}[ht]
	\centering
	\includegraphics[width=1 \columnwidth]{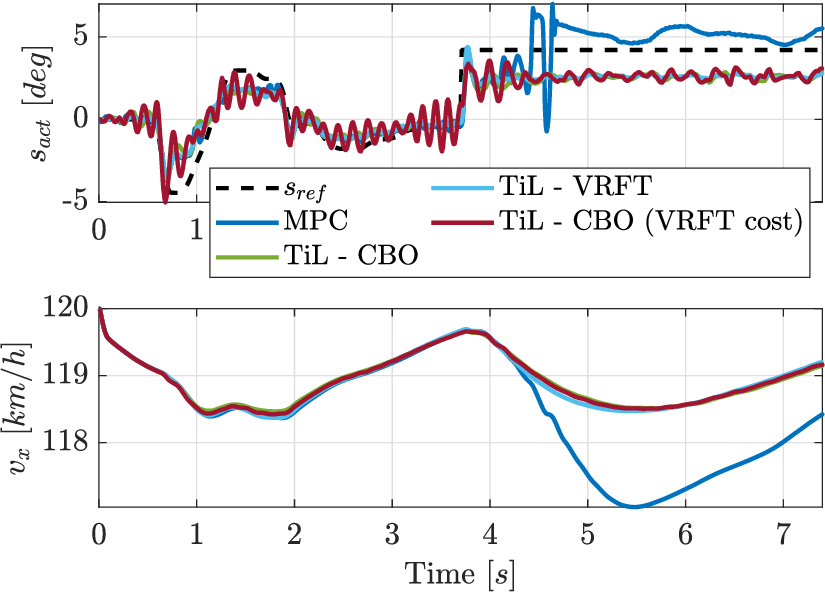}
	\caption{Controller validation onto the optimization experiment: speed and steer profiles.}
	\label{fig:til_opt_optimization_experiment_speed_steer}
\end{figure}

\begin{table}
	\resizebox{\columnwidth}{!}{
		\begin{tabular}{llll}
			\hline
			\textbf{Controller} & $rms(\tilde{r}-r)$ $\left[deg/s\right]$ & $rms(\tilde{\beta}-\beta)$ $\left[deg\right]$ & $rms(\dot{s}_{act})$ $\left[deg/s\right]$ \\ \hline
			MPC                 & $1.65$                                 & $1.46$                                     & $921.92$ \\ \hline
			TiL-SMGO-$\Delta$              & $0.75$                                 & $1.07$                                     & $ 462.35$ \\ \hline
			TiL-VRFT            & $1.17$                                 & $1.25$                                     & $424.56$ \\ \hline
		\end{tabular}
	}
	\caption{Performance of the controllers in the first validation test (double-lane-change with step-steer at $140\ km/h$).}
	\label{tab:til_opt_testing_test_2_metrics}
\end{table}

We show here the time-domain results when considering the optimization test, \emph{i.e.} the double lane change followed by a step steer, at $120\ km/h$. Figure \ref{fig:til_opt_optimization_experiment_tracking} depicts the vehicle yaw-rate (upper plot) and sideslip angle (bottom plot). The black dashed line $\tilde{r}$ is the reference behaviour obtained in nominal conditions by using MPC onto the digital twin (see the scheme of Fig. \ref{fig:til_opt_silc_scheme}). The nominal MPC exhibits some issues in the final part of the experiment, and fails at maintaining the sideslip angle at an acceptable level, while also badly tracking the yaw-rate. On the other hand, the TiL enhanced controller is capable of smoothly recovering the system behaviour and guaranteeing good yaw-rate tracking and minimization of the sideslip. Note that perfectly ''tracking'' the sideslip is not possible (nor it is required), since the cornering stiffness is different and the physical vehicle has a different steady-state response. 

Looking at the performance obtained with VRFT in the time-domain confirms that the method is adequate for learning a TiL controller, and significantly eases the tuning phase, at the cost of a minimum performance reduction (note that VRFT is achieving in a single iteration what BO or SMGO are able to do only after 10 or 15 experiments).
When considering SMGO with VRFT cost (\emph{i.e.} filtering the output errors within the cost function), we note an oscillating response. This is explainable, as the pre-filtering applied onto the measured signals during the training process ''hides'' some high- frequencies, and SMGO is in fact over-fitting the controller onto a filtered version of the system, poorly performing outside of that frequency range.

For the same experiment, Fig. \ref{fig:til_opt_optimization_experiment_speed_steer} depicts speed (constant) and steer profiles when using the same controllers. What observed from the oscillating yaw-rate response in the SMGO with VRFT cost is further noted by the steer profile, which is particularly aggressive. $s_{ref}$ is the driver steer request: one can note as the controllers cut the request and apply a slightly different steering angle, so as to guarantee the vehicle stability and avoid spinning (\emph{i.e.} too high sideslip angles).
Finally, Table \ref{tab:til_opt_optimization_test_metrics} quantifies the cost indexes for the given experiment and the controllers.
\subsubsection{Double lane change with step steer - $140\ km/h$}
\begin{figure}[ht]
	\centering
	\includegraphics[width=1 \columnwidth]{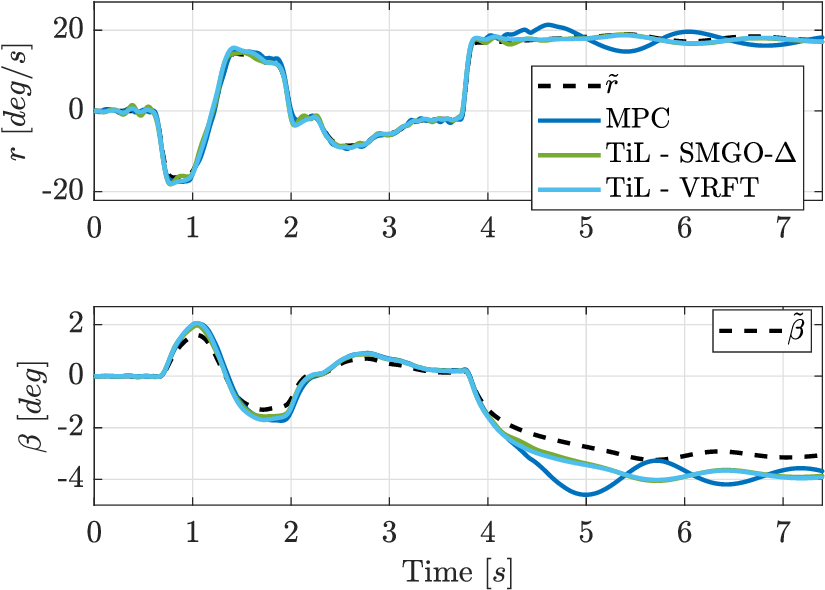}
	\caption{Controller validation onto the first testing experiment: yaw-rate and sideslip tracking.}
	\label{fig:til_opt_testing_experiment_2_tracking}
\end{figure}
\begin{figure}[ht]
	\centering
	\includegraphics[width=1 \columnwidth]{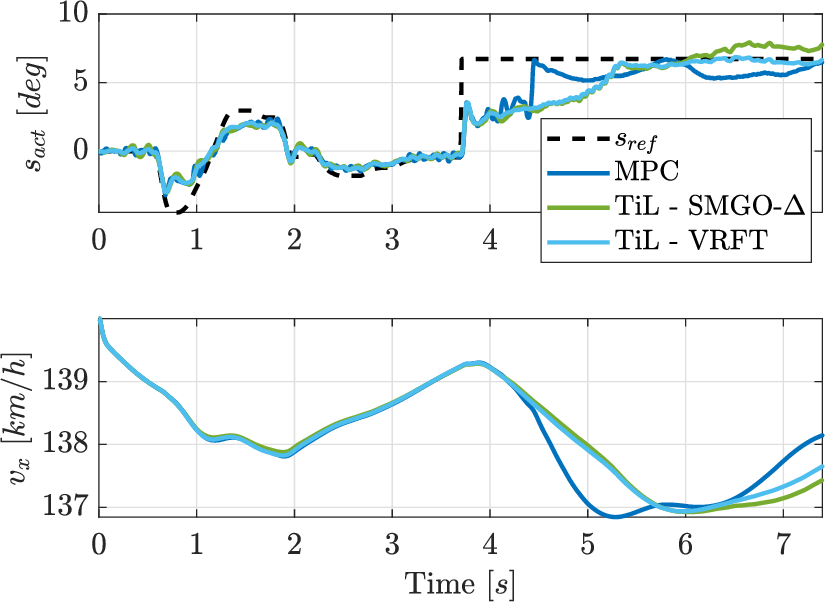}
	\caption{Controller validation onto the first testing experiment: speed and steer profiles.}
	\label{fig:til_opt_testing_experiment_2_speed_steer}
\end{figure}
\begin{figure}[ht]
	\centering
	\includegraphics[width=1 \columnwidth]{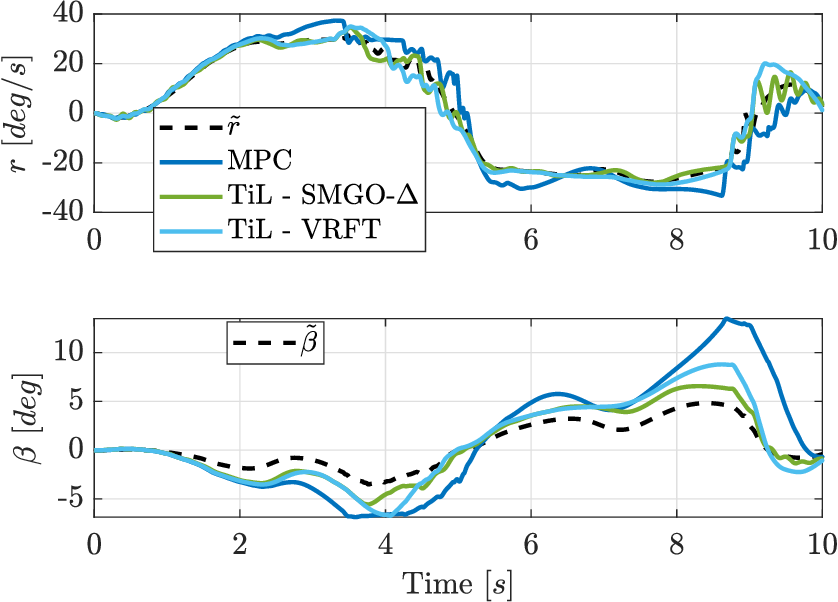}
	\caption{Controller validation onto the second testing experiment: yaw-rate and sideslip tracking.}
	\label{fig:til_opt_testing_experiment_tracking}
\end{figure}
\begin{figure}[ht]
	\centering
	\includegraphics[width=1 \columnwidth]{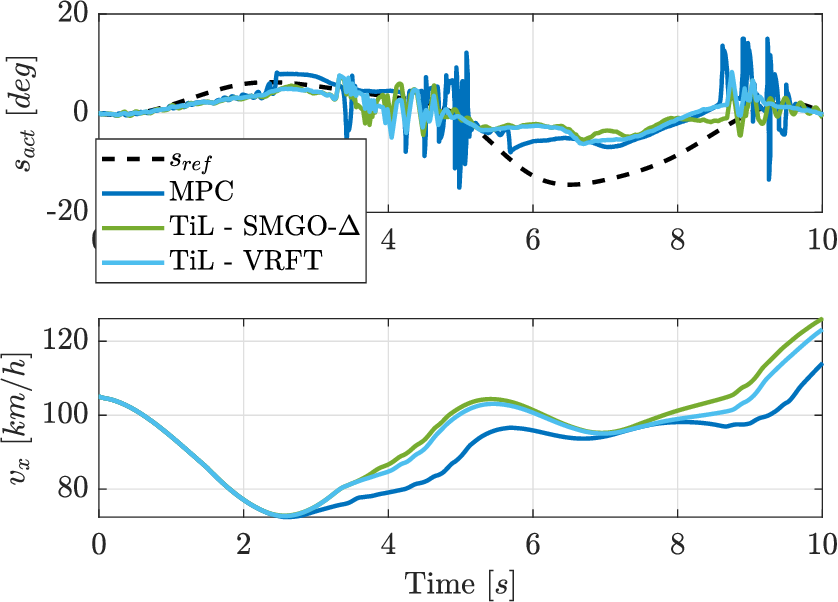}
	\caption{Controller validation onto the second testing experiment: speed and steer profiles.}
	\label{fig:til_opt_testing_experiment_speed_steer}
\end{figure}

We show here the time-domain results when considering a double lane change followed by a step steer, at $140\ km/h$. Figure \ref{fig:til_opt_testing_experiment_tracking} depicts the vehicle yaw-rate (upper plot) and sideslip angle (bottom plot). As in the previous case, TiL is able to make the real vehicle track the digital twin: both SMGO and VRFT calibrations suffices for our purposes, and yield very similar results. Figure \ref{fig:til_opt_testing_experiment_speed_steer} shows the speed and steer profiles in the same test, while Table \ref{tab:til_opt_testing_test_1_metrics} provides some quantitative metrics.
\subsubsection{Chikane test}

\begin{table}
	\resizebox{\columnwidth}{!}{
		\begin{tabular}{llll}
			\hline
			\textbf{Controller} & $rms(\tilde{r}-r)$ $\left[deg/s\right]$ & $rms(\tilde{\beta}-\beta)$ $\left[deg\right]$ & $rms(\dot{s}_{act})$ $\left[deg/s\right]$ \\ \hline
			MPC                 & $5.40$                                 & $3.73$                                     & $3.09 \times 10^3$ \\ \hline
			TiL-SMGO-$\Delta$               & $2.30$                                 & $1.25$                                     & $1.39 \times 10^3$ \\ \hline
			TiL-VRFT            & $3.59$                                 & $1.92$                                     & $1.19 \times 10^3$ \\ \hline
		\end{tabular}
	}
	\caption{Performance of the controllers in the second validation test (chikane)}
	\label{tab:til_opt_testing_test_1_metrics}
\end{table}

We show here the time-domain results when considering the second validation test, \emph{i.e.} a chikane maneuver inspired from the Savelli curve (nr. 7 in Mugello track). In this case, a realistic speed profile is considered (displayed in the bottom plot of Fig. \ref{fig:til_opt_testing_experiment_2_speed_steer}), with the driver pushing the gas pedal at the middle of the curve, thus combining lateral with longitudinal acceleration; this kind of profile if particularly challenging for the considered MPC controller, which assumes the longitudinal speed is constant in the prediction horizon. In fact, from Fig. \ref{fig:til_opt_testing_experiment_2_tracking}, one can note the worsened yaw-rate tracking performance, with respect to constant speed tests. Also in this case, the nominal MPC exhibits a much worse behaviour than TiL-enhanced ones. This is further noted from the steering command, in the upper plot of Fig. \ref{fig:til_opt_testing_experiment_2_speed_steer}, which shows significant oscillations. Table \ref{tab:til_opt_testing_test_2_metrics} gives the quantitative metrics for the chikane test: it is interesting to note that although slightly less performing, VRFT calibration is somewhat less aggressive in terms of $rms(\dot{s}_{act})$.

\section{Conclusions}
\label{Section:til_opt_Conclusions}
In this paper, we took a challenging case study for the recently proposed TiL-C architecture, namely, the yaw-rate tracking problem, for which a baseline Model Predictive Controller is designed, and then enhanced with a simple PID compensator. A thorough discussion on different tools for optimizing the compensator gains is carried out: specifically, we compared the widely popular Bayesian Optimization with the newly proposed Set Membership Global Optimization. We also consider a philosophically different approach, the Virtual Reference Feedback Tuning, which is a direct control design method, requiring just one batch of open-loop data for obtaining the compensator gains.

The comparisons show as VRFT is capable of getting very close to the global minimum, with just one experiment, while BO and SMGO need at least $10-15$ tests to overcome it, and fail at converging at the best point if not provided with the previously computed VRFT controller gains.

The time-domain results validate the calibrated controllers in different scenarios, considering both constant and varying speed profiles, thus confirming the effectiveness of TiL-C with respect to the baseline controller, and showing the amount of improvement achieved by performing some iterations of BO/SMGO, as compared to VRFT.

\section*{Acknowledgments}
We would like to thank companies VI-Grade GmbH and Ferrari S.p.A. for the technical support provided during the work.
\bibliographystyle{elsarticle-harv}
\bibliography{ref}

\end{document}